\documentclass[twoside,leqno,twocolumn]{article}

\usepackage[algo2e,ruled,vlined,linesnumbered]{algorithm2e}
\usepackage{ltexpprt}

\usepackage{hyperref}
\usepackage{booktabs}
\usepackage{multirow}
\usepackage{amsfonts}
\usepackage[table]{xcolor}
\usepackage{todonotes}
\usepackage{subcaption}
\usepackage{siunitx}
\usepackage{microtype}
\usepackage{xspace}

\graphicspath{{./graphics/}}%helpful if your graphic files are in another directory
\begin{document}

\title{\Large Efficiently Computing Maximum Flows in Scale-Free Networks}
%\thanks{Supported by GSF grants ABC123, DEF456, and GHI789.}}
\author{Thomas Bl\"asius \and Tobias Friedrich \and Christopher Weyand}
%\and Tricia Manning\thanks{Society for Industrial and Applied Mathematics.}}

\date{}

\maketitle

% Copyright Statement
% When submitting your final paper to a SIAM proceedings, it is requested that you include 
% the appropriate copyright in the footer of the paper.  The copyright added should be 
% consistent with the copyright selected on the copyright form submitted with the paper.
% Please note that "20XX" should be changed to the year of the meeting.

% Default Copyright Statement
%\fancyfoot[R]{\scriptsize{Copyright \textcopyright\ 2021 by SIAM\\
%Unauthorized reproduction of this article is prohibited}}

% Depending on which copyright you agree to when you sign the copyright form, the copyright 
% can be changed to one of the following after commenting out the default copyright statement
% above.

%\fancyfoot[R]{\scriptsize{Copyright \textcopyright\ 20XX\\
%Copyright for this paper is retained by authors}}

%\fancyfoot[R]{\scriptsize{Copyright \textcopyright\ 20XX\\
%Copyright retained by principal author's organization}}

%\pagenumbering{arabic}
%\setcounter{page}{1}%Leave this line commented out.

\begin{abstract} \small\baselineskip=9pt 
We study the maximum-flow/minimum-cut problem on scale-free networks, i.e., graphs whose degree distribution follows a power-law.
We propose a simple algorithm that capitalizes on the fact that often only a small fraction of such a network is relevant for the flow.
At its core, our algorithm augments Dinitz's algorithm with a balanced bidirectional search.
Our experiments on a scale-free random network model indicate sublinear run time.
On scale-free real-world networks, we outperform the commonly used highest-label Push-Relabel implementation by up to two orders of magnitude.
Compared to Dinitz's original algorithm, our modifications reduce the search space, e.g., by a factor of 275  on an autonomous systems graph.

Beyond these good run times, our algorithm has an additional advantage compared to Push-Relabel.
The latter computes a preflow, which makes the extraction of a minimum cut potentially more difficult.
This is relevant, for example, for the computation of Gomory-Hu trees.
On a social network with \SI{70000} nodes, our algorithm computes the Gomory-Hu tree in 3 seconds compared to 12 minutes when using Push-Relabel.

\end{abstract}

\section{Introduction}

The maximum flow problem is arguably one of the most fundamental graph
problems that regularly appears as a subtask in various applications
\cite{Ahuja1993, Schaeffer2007, Verma2012}.  The go-to general-purpose
algorithm for computing flows in practice is the highest-label
Push-Relabel algorithm by Cherkassky and
Goldberg~\cite{Cherkassky1997}, which is also part of the boost graph
library~\cite{schaeling2011}.  Beyond that, the BK-algorithm by Boykov
and Kolmogorov~\cite{Boykov2004} or its later
iteration~\cite{Goldberg2011} should be used for instances
appearing in computer vision.  Our main goal in this paper is to
provide a flow algorithm tailored towards \emph{scale-free} networks.
Such networks are characterized by their heavy-tailed degree
distribution resembling a power-law, i.e., they are sparse with few
vertices of comparatively high degree and many vertices of low degree.

At its core, our algorithm is a variant of Dinitz's algorithm~\cite{Dinitz1970}.
Dinitz's algorithm is an augmenting path algorithm that
iteratively increases the flow along collections of shortest paths in the residual network.  In each
iteration, at least one edge on every shortest path gets saturated,
thereby increasing the distance between source and sink in the
residual network.  To exploit the structure of scale-free networks, we
make use of the facts that, firstly, shortest paths tend to span only
a small fraction of such networks, and secondly, a balanced
bidirectional breadth first search is able to find the shortest paths
very efficiently~\cite{Borassi2016,Blaesius2018}.  Using a
bidirectional search to compute the collection of shortest paths in
Dinitz's algorithm directly translates this efficiency to the first
iteration, as the residual network initially coincides with the flow
network.  Though the structure of the residual network changes in
later iterations, our experiments show that the run time improvements
achieved by using a bidirectional search remain high.

Scaling experiments with geometric inhomogeneous random graphs
(GIRGs)\footnote{GIRGs are a generative network model closely related
  to hyperbolic random graphs~\cite{Krioukov2010}.  They resemble real-world networks in
  regards to important properties such as degree distribution,
  clustering, and distances.}~\cite{Bringmann2019}, in fact indicate
that the flow computation of our algorithm runs in sublinear time.  In
comparison, previous algorithms (Push-Relabel, BK, and unidirectional
Dinitz) require slightly super-linear time.  This is also reflected in
the high speedups we achieve on real-world scale-free networks.

With the flow computation itself being so efficient, the total run
time for computing the maximum flow for a single source-sink pair in
a scale-free network is heavily dominated by loading the graph and
building data structures.  Thus, our
algorithm is particularly relevant when we have to compute multiple
flows in the same network.  This is, e.g., the case when computing the
Gomory-Hu tree~\cite{Gomory1961} of a network.  The Gomory-Hu tree is
a compact representation of the minimum $s$-$t$ cuts for all source-sink pairs $(s, t)$.  It can be
computed with Gusfield's algorithm~\cite{Gusfield1990} using $n - 1$
flow computations in a network with $n$ vertices.  Using our
bidirectional flow algorithm as the subroutine for flow computations
in Gusfield's algorithm lets us compute the Gomory-Hu tree of, e.g.,
the \texttt{soc-slashdot} instance with \SI{70}{\kilo\relax} nodes and
\SI{360}{\kilo\relax} edges in only \SI{2.6}{\second}.  In this context, we
observe that the Push-Relabel algorithm is also very efficient in
computing the flow values by computing a preflow.  However, converting
this to a flow or extracting a cut from it takes significantly more
time.

Our algorithm is designed to work particularly
well on scale-free networks.  Nonetheless, we also conducted
experiments on networks that are not scale-free.  We observe that our
algorithm outperforms the Push-Relabel algorithm significantly on
Erdős-Rényi random graphs and slightly on the Pennsylvania road
network.  Unsurprisingly, our algorithm is outperformed by the
BK-algorithm on a segmentation instance from computer vision.
Moreover, Push-Relabel performs best on a layered network that was
specifically constructed to evaluate flow algorithms.  However, we
would argue that this type of instance is rather artificial.

\subsection{Contribution.}

Our findings can be summarized in the following main contributions.
\begin{itemize}
\item We provide a simple and efficient flow-algorithm that
  significantly outperforms previous algorithms on scale-free
  networks.
\item It's efficiency on non-scale-free instances makes it a potential
  replacement for the Push-Relabel algorithm for general-purpose flow
  computations.
\item Our algorithm is well suited to compute the Gomory-Hu tree of
  comparatively large instances.
\item There are situations, where computing a flow with the
  Push-Relabel algorithm is significantly more expensive than
  computing a preflow.  This stands in contrast to previous
  observations~\cite{Cherkassky1997, Derigs1989}.
\end{itemize}

\subsection{Related Work.}
\label{sec:related-work}

The maximum flow problem has been for a long time and still is subject
of active research. In the following,
we briefly discuss only the work most related to our result.  For a
more extensive overview on the topic of flows, we refer to the survey
by Goldberg and Tarjan~\cite{gt-emfa-14}.

Our algorithm is based on Dinitz's Algorithm~\cite{Dinitz1970}, which belongs to the
family of \emph{augmenting path algorithms} originating from the
Ford-Fulkerson algorithm~\cite{Ford1956}.  Augmenting path algorithms
use the \emph{residual network} to represent the remaining capacities
and iteratively increase the flow by augmenting it with paths from
source to sink in the residual network, until no such path exists.  At
every point in time, a valid flow is known and at the end of
execution, non-reachability in the residual network certifies
maximality.

From this perspective, the \emph{Push-Relabel
  algorithm}~\cite{Goldberg1988} does the reverse.  At every point in
time, the sink is not reachable from the source in the residual
network, thereby guaranteeing maximality, while the object maintained
throughout the algorithm is a so-called \emph{preflow} and the
algorithm stops once the preflow is actually a flow.  This is achieved
using two operations \emph{push} and \emph{relabel}; hence the name.
Different variants of the Push-Relabel algorithm mainly
differ with regards to the order in which operations are applied.  
A strategy performing well in practice is the
highest-label strategy~\cite{Cherkassky1997}.  The extensive empirical
study by Ahuja et al.~\cite{Ahuja1997} on ten different
algorithms shows that the highest-label Push-Relabel algorithm indeed
performs the best out of the ten.  The only small caveat with these experiments is the
fact that they are based on artificial networks that are specifically
generated to pose difficult instances.  Our experiments show that the
structure of the instance matters in the sense that it impacts
different algorithms differently; potentially yielding different
rankings on different types of instances.
The so-called pseudoflow algorithm by Hochbaum~\cite{Hochbaum2008} was
later shown to slightly outperform (low single-digit speedups on most
instances) the highest-label Push-Relabel algorithm; again based on
artificial instances~\cite{Chandran2009}.

Boykov and Kolmogorov~\cite{Boykov2004} gave an algorithm tailored
specifically towards instances that appear in computer vision;
outperforming Push-Relabel on these instances.  It was later refined
by Goldberg et al.~\cite{Goldberg2011}.  Most related to our studies
is the work by Halim et al.~\cite{Halim2011} who developed a
distributed flow algorithm for MapReduce to compute flows on huge
social networks.

\section{Network Flows and Dinitz's Algorithm}

In this section we introduce the concept of network flow and describe Dinitz's algorithm \cite{Dinitz1970}.

% flow
\paragraph{Network Flows.}
A flow network is a directed graph $G=(V,E)$ with source and sink vertices $s,t\in V$, and a capacity function $c: V\times V\to \mathbb{N}$ with $c(u,v)=0$ if $(u,v)\not\in E$.
A \emph{flow} $f$ on $G$ is a function over vertex pairs $f: V\times V \to \mathbb{Z}$ satisfying three constrains: 
(I) capacity $f(u,v) \leq c(u,v)$
(II) asymmetry $f(u,v) = - f(v,u)$ and
(III) conservation $\sum_{v\in V} f(u,v) = 0$ for $u\in V\setminus \{s,t\}$.
We call an edge $(u,v)\in E$ \emph{saturated} if $f(u,v)=c(u,v)$.
Denote the \emph{value} of a flow $f$ as $\sum_{v\in V}f(s,v)$.
The maximum flow problem, \emph{max-flow} for short, is the problem of finding a flow of maximum value.

Given a flow $f$ in $G$, we define a network $G_f$ called the \emph{residual network}.
$G_f$ has the same set of nodes and contains the directed edge $(u,v)$ if $f(u,v) < c(u,v)$.
The capacity $c'$ of edges in $G_f$ is given by the residual capacity in the original network, i.e., $c'(u,v) = c(u,v) - f(u,v)$.
An $s$-$t$ path in $G_f$ is called an \emph{augmenting path}.

% res net

% dinics description
\paragraph{Dinitz's Algorithm.}
One can solve max-flow by iteratively increasing the flow on augmenting paths, yielding the famous Ford-Fulkerson algorithm~\cite{Ford1956}.
Dinitz's algorithm belongs to the family of augmenting path algorithms \cite{Ahuja1993}.
In contrast to the Ford-Fulkerson algorithm, Dinitz groups augmentations into rounds.

% layered network
Let $d_s(v)$ be the distance from $s$ to vertex $v$ in $G_f$. 
We define a subgraph of $G_f$ called the \emph{layered network} by restricting the edge set to edges $(u,v)$ of $G_f$ for which $d_s(u)+1 = d_s(v)$, i.e., edges that increase the distance to the source.
% blocking flow
We call a flow of some network a \emph{blocking flow} if every $s$-$t$ path contains at least one edge that is saturated by this flow, i.e.,~there is no augmenting path.

Each round, Dinitz's algorithm (see Algorithm~\ref{alg:dinics})
augments a set of edges that constitutes a blocking flow of the layered network.
One can find such a set of edges by iteratively augmenting $s$-$t$ paths in the layered network until source and sink become disconnected.
After augmenting a blocking flow, the distance between the terminals in the residual network strictly increases.
%Since each $s$-$t$ path in the layered network is a shortest $s$-$t$ path in the residual network, Dinitz can be understood as augmenting a maximal set of shortest paths in the residual graph each round.

\begin{algorithm2e}
\DontPrintSemicolon
\While(){s-t path in residual network}{
    build layered network\;
    \While(){s-t path in layered network}{    
        augment flow with s-t path
    }
}
%\While(\tcc*[l]{at most $n-1$ times}){s-t path in residual network}{
%    build layered network                   \tcc*{$O(m)$ using BFS}
%    \While(\tcc*[f]{at most $m$ times}){s-t path in layered network}{    
%        augment flow with s-t path          \tcc*{amortized $O(n)$ using DFS}
%    }
%}
\caption{Dinitz's Algorithm.}
\label{alg:dinics}
\end{algorithm2e}

% runtime
\paragraph{Asymptotic Running Time.}
%To describe and evaluate our modifications later, we outline the runtime analysis.
To better understand how our modifications impact the run time, we briefly sketch how Dinitz running time of $O(n^2m)$ is obtained.
Since $d_s(t)$ increases each round, the number of rounds is bounded by $n-1$.
Each round consists of two stages: building the layered network and augmenting a blocking flow.
To build the layered network, the distances from the source to every vertex in the residual network are needed.
The layered network can be constructed in $O(m)$ using a breadth-first search (BFS).
Asymptotically, however, this is dominated by the time to find the blocking flow.
Finding the paths of the blocking flow is done with a repeated graph traversal, usually using a depth-first search (DFS). 
%A path can be augmented as soon as it is found.
%Edges that are introduced into the residual network due to augmentation will never be part of the current layered network as they are directed from sink to source.
%Edges that become saturated can be deleted from the layered network, because the blocking flow is only required to be maximal and not maximum.
The number of found paths is bounded by $m$, because each found path saturates at least one edge, removing it from the layered network.
A single DFS can be done in amortized $O(n)$ time as follows.
Edges that are not part of an $s$-$t$ path in the layered network do not need to be looked at more than once during one round.
This is achieved by remembering for each node which edges of the layered network were already found to have no remaining path to the sink.
Each subsequent DFS will start where the last one left off.
Thus, per round, the depth-first searches have a combined search space of $O(m)$, while each individual search additionally visits the nodes on one $s$-$t$ path which is $O(n)$.

\paragraph{Efficient Dinitz Implementation.}
Typical implementations represent the graph by adding a reversed twin for each edge.
Furthermore, neither the residual network nor the layered network are constructed explicitly.
The residual network is implicitly defined by the capacities and flow values on edges and the layered network by a distance labeling.
This conveniently eliminates the need to modify the network structure during the algorithm.
When, e.g., saturating an edge during augmentation, this implicitly removes the edge from the residual network and layered network.
However, with this representation, the BFS and DFS are performed on all edges and must check if edges are part of the residual or layered network when they are encountered.
The bound for the BFS is unaffected and the amortization argument for the DFS extends to edges that are not part of the layered and/or residual network.
During the augmentation of the blocking flow, a counter into the adjacency list of each vertex indicates which outgoing edges were already processed this round.

\paragraph{Practical Performance.}
The practical performance of Dinitz's algorithm is far better than its worst-case bound.
Actually, $O(n)$ as the length of the found augmenting path is very unrealistic.
In our experiments $d_s(t)$ remains mostly below 10, implying that the number of rounds is significantly lower than $n-1$.
Also, the number of found augmenting paths during one rounds is far below $O(m)$.
In unweighted networks, for example, a DFS saturates all edges of the found path resulting in a bound of $O(m)$ to find a blocking flow.
In fact, Dinitz's algorithm has a tight upper bound of $O(n^{2/3}m)$ in unweighted networks \cite{Even1975, Karzanov1973}.

%\clearpage

% 
% Optimizing and Implementing Dincics
%     Bidir search space with pictures
%     5 optimizations (just state)
% 

\section{Improving Dinitz on Scale-Free Networks}
\label{sec:opt}

We adapt a common Dinitz implementation\footnote{\url{https://cp-algorithms.com/graph/dinic.html}} to exploit the specific structure of scale-free networks.
We achieve a significant speedup by using the fact that a flow and cut respectively often depend only on a small fraction of the network.
%Flow algorithms are very efficient in practice \cite{Boykov2004, Lang2004, Verma2012}.
The following three modifications each tackle a performance bottleneck.

\begin{figure}[t]
\centering
\includegraphics[width=0.9\columnwidth]{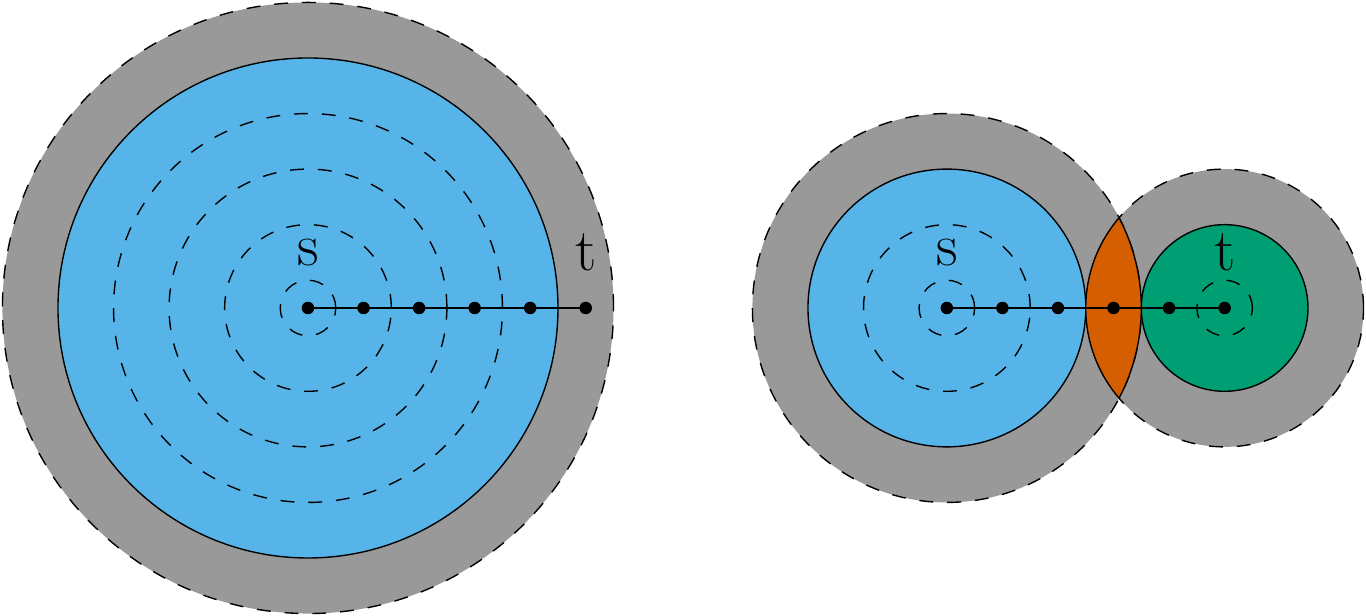}
\caption{Search space of a breadth-first search from a source $s$ to a sink $t$ unidirectional (left) and bidirectional (right).
The blue area represents the vertices that are explored, i.e., whose outgoing edges were scanned, by the forward search and the green area the backward search.
In the gray area are vertices that are seen during exploration of the last layer, but not yet explored.
Vertices in the intersection of the upcoming layers of the backward and forward search are marked orange.
%The unidirectional search ends when $t$ is encountered.
%The bidirectional search ends when a vertex is encountered that was already seen from the other direction, here indicated in orange.
}
\label{fig:opt-space}
\end{figure}

\paragraph{Bidirectional Search.}
Recently, sublinear running time was shown for balanced bidirectional search in a scale-free network model \cite{Blaesius2018, Borassi2016}.
We use a bidirectional breadth-first-search to compute the distances that define the layered network during each round of Dinitz's algorithm.
%The theoretical bounds regarding the bidirectional search only hold for the very first round, because the residual network changes during the algorithm.
%Nevertheless we assume the bidirectional search to also reduce the search space in subsequent rounds.
A forward search is performed from the source and a backward search from the sink, each time advancing the search that incurs the lower cost to advance one layer.
A shortest $s$-$t$ path is found when a vertex is discovered that was already seen from the other direction.
Note that, for our purpose, the bidirectional search has to finish the current layer when such a vertex is discovered, because all shortest paths must be found.
Figure~\ref{fig:opt-space} visualizes the difference in explored vertices between a normal and a bidirectional BFS.
The augmentations with DFS are restricted to the visited part of the layered network, meaning the search space of the BFS plus the next layer.

The distance labeling obtained by the bidirectional BFS requires a change to the DFS.
The purpose of the layered network is to contain all edges on shortest $s$-$t$ paths.
The DFS identifies edges $(u,v)$ of the layered network by checking if they increase the distance from the source, i.e.,~$d_s(u)+1=d_s(v)$.
However, we no longer obtain the distances from the source for all relevant vertices.
For vertices processed by the backward search, distances to the sink $d_t(v)$ are known instead.
To resolve the problem, we allow edges that either increase distance from the source or decrease distance to the sink, i.e.,~$d_s(u)+1=d_s(v)$ or $d_t(u)-1=d_t(v)$.
This deviates from the definition of the layered network.
But since edges on shortest $s$-$t$ paths must both, increase the distance from the source and decrease the distance to the sink, we do not miss any relevant edges.
%Furthermore, this definition has the additional advantage that the DFS cannot deviate from shortest $s$-$t$ paths in the search space of the backward search.
%First, the search space of the backward search can only be entered by the DFS via vertices on a shortest path.
%Second, when already on a shortest path, each edge that decreases the distance to the sink will also be on a shortest path.

\paragraph{Time Stamps.}
% what is bottleneck now? -> init
The bidirectional search reduces the search space of the breadth-first search and depth-first search substantially, potentially to sublinear.
The initialization, however, still requires linear time.
% what has to be done in init?
It includes the following.
For the BFS, distances from the source and to the sink must be initialized to infinity.
For the augmentations, one counter per node has to be initialized to zero.

% how do we make it fast?
To avoid the linear initializations, we introduce time stamps to indicate if a vertex was seen during the current round.
The initialization of distances and counters is done lazily as vertices are discovered during the BFS.
Another detail of our implementation is that we use begin and end indices into an array instead of a dynamically growing queue for the BFS.
We allocate this memory in advance and override the data each round.

\paragraph{Skip Next Forward Layer.}
Recall that we identify edges of the layered network by checking if they increase the distance from the source or decrease the distance to the sink.
Therefore the DFS proceeds along edges outgoing from the last forward search layer independent from the target vertex being seen only by the forward search (gray in Figure~\ref{fig:opt-space}) or also by the backward search (orange in Figure~\ref{fig:opt-space}).
However, the former type of vertex cannot be part of a shortest $s$-$t$ path.
By saving the number of explored layers of the forward search we can avoid the exploration of such vertices, thus limiting the DFS to vertices colored blue, green, or orange in Figure~\ref{fig:opt-space}.
With this optimization, the combined search space during augmentation (lines 3,4 in Algorithm~\ref{alg:dinics}) is almost limited to the search space of the BFS.
The only additional edges that are visited originate from the intersection of the forward and backward search.
%Lastly, edges can be visited multiple times in weighted networks if they are on more than one shortest s-t path.

%\clearpage

% 
% Evaluation of OPTs on Scale-Free networks
% 

\section{Experimental Evaluation}

% general remarks
In this section, we investigate the performance of our algorithm \emph{DinitzOPT}\footnote{The code will be available upon publication.}.
First, we compare it to established approaches on real-world networks in Section~\ref{sec:eval-runtime}.
We additionally examine the scaling behavior and how the comparison is affected by problem size,~i.e., is there an asymptotic improvement over other algorithms?
Then, Section~\ref{sec:eval-detail} evaluates to which extent the different optimizations contribute to better run times and search space.
In Section~\ref{sec:eval-gh} we analyze the algorithms in a specific application (Gomory-Hu trees) and compare their usability beyond the speed of the actual flow computation.
To this end, we test three different approaches to obtain a cut with the Push-Relabel algorithm.
Lastly, we extent our considerations to other types of networks in Section~\ref{sec:eval-other} and discuss why the results on scale-free networks differ from previous studies.
Recall that bidirectional search was found to perform particularly well on heterogeneous networks.

\subsection{Runtime Comparison.}
\label{sec:eval-runtime}

\begin{figure*}
\centering
\includegraphics[width=\textwidth]{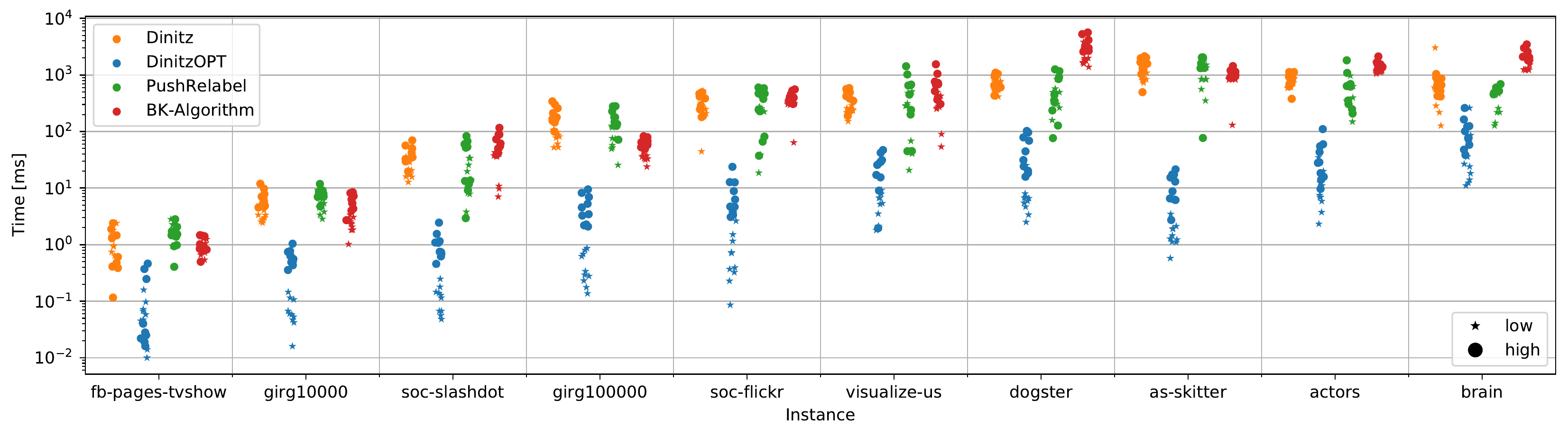}
\caption{Runtime comparison of flow computations.
The 20 computed flows per instance are divided into \emph{low} and \emph{high} terminal pairs.
For \emph{low}, the terminal degree is between 0.75 and 1.25 times the average degree. 
For \emph{high}, it is between 10 and 100 times the average degree.
Pairs are chosen uniformly at random from all vertices with the respective degree.
}
\label{fig:eval-runtime}
\end{figure*}

% setup talk
In this section we compare our new approach to three existing algorithms: Dinitz~\cite{Dinitz1970}, Push-Relabel~\cite{Goldberg1988}, and the Boykov-Kolmogorov (BK) algorithm~\cite{Boykov2004}.
We modified their respective implementations to support our experiments.
This also includes some minor performance-relevant changes listed in the appendix (see Section~\ref{sec:impl}).
The experiments include two synthetic and eight real-world networks.
All networks are undirected and all but \texttt{visualize-us} and \texttt{actors} are unweighted.
Further details regarding the datasets can be found in Table~\ref{tab:instances}.
We restrict our experiments in this section to the flow computation only.
That is, the measurements exclude the time it takes to initialize intermediate data structures before and after flow computations as well as the creation of the graph structure.
For Push-Relabel we only measure the computation of the \emph{preflow}, which is sufficient to determine the value of the flow/cut.
Figure~\ref{fig:eval-runtime} shows the resulting run times.
For this plot, the terminals were chosen uniformly at random from the set of vertices with degree close to the average (\emph{low}) or considerably higher degree (\emph{high}).

% general observations
One can see that Dinitz and Push-Relabel display comparable times while BK is slightly slower on most large instances.
DinitzOPT consistently outperforms the other algorithms by one to three orders of magnitude.
% opt is more problem sensitive
The variance is also higher for DinitzOPT with \emph{low} pairs approximately one order of magnitude faster on average than \emph{high} pairs. 
This is best seen in the \texttt{girg100000} instance and suggests that DinitzOPT is able to better exploit easy problem instances.
For all other algorithms the effect of the terminal degree on the run time is barely noticeable.
% random pairs are too easy? -> cuts are often around one terminal
Another observation is that all algorithms display drastically lower run times than their respective worst-case bounds would suggest.

% easy instances
The times in our experiments are close to what one might expect from linear algorithms.
For example, Dinitz computes a flow on the \texttt{as-skitter} instance in one second.
Considering the tight $O(mn^{2/3})$ bound in unweighted networks and assuming the throughput per second to be around $10^8$ --- which is a generous guess for graph algorithms --- would result in an estimate of 30 minutes per flow.
In this context, there are also experimental results that appear to conflict with our results.
Earlier studies found Dinitz to be slower than Push-Relabel and both algorithms clearly super-linear on a series of synthetic instances~\cite{Ahuja1997}.
However, these synthetic instances exhibit specifically crafted hard structures that are placed between designated source and sink vertices.
These instances thus present substantially more challenging flow problems.
We assume the low times in our experiments to be caused by the scale-free network structure and, to a lesser degree, the simplicity of the problem instances when choosing a random pair of nodes as terminals.
Furthermore, most of our instances are unweighted and undirected.

% terminal degree and trivial cuts
\paragraph{Effect of the Terminal Degree.}
In the following, we discuss the effect of terminal degree and structure of the cut on the run time of Dinitz and DinitzOPT.
Note that the terminal degree is an upper bound on the size of the cut in unweighted networks. 
Moreover, the terminal degree in our experiments is based on the average degree, which is assumed to be constant in many real-world networks~\cite{Barabasi2016}.
Thus, the $O(mC)$ bound for augmenting path based algorithms, with $C$ being the size of the cut, implies not only a linear bound for the eight unweighted networks in our experiments, but would also explain faster \emph{low} pairs.
Surprisingly, DinitzOPT exploits low terminal degrees much more than Dinitz.
Another explanation for faster \emph{low} pairs is that many cuts are close around one terminal, which is consistent with previous observations about cuts in scale-free networks~\cite{Leskovec2009,Son2006}.
Moreover, Dinitz tends to perform well when the source side of the cut is small~\cite{Orecchia2014}.
Although this does not fully explain why DinitzOPT is more sensitive to the terminal degree, we observe in Section~\ref{sec:eval-gh} that Dinitz slows down massively when the source degree is high, even with low sink degree.
Since DinitzOPT always advances the side with smaller volume during bidirectional search it does not matter which terminal has the higher degree.

% scaling
\begin{figure}[t]
    \centering
    \includegraphics[width=0.9\columnwidth]{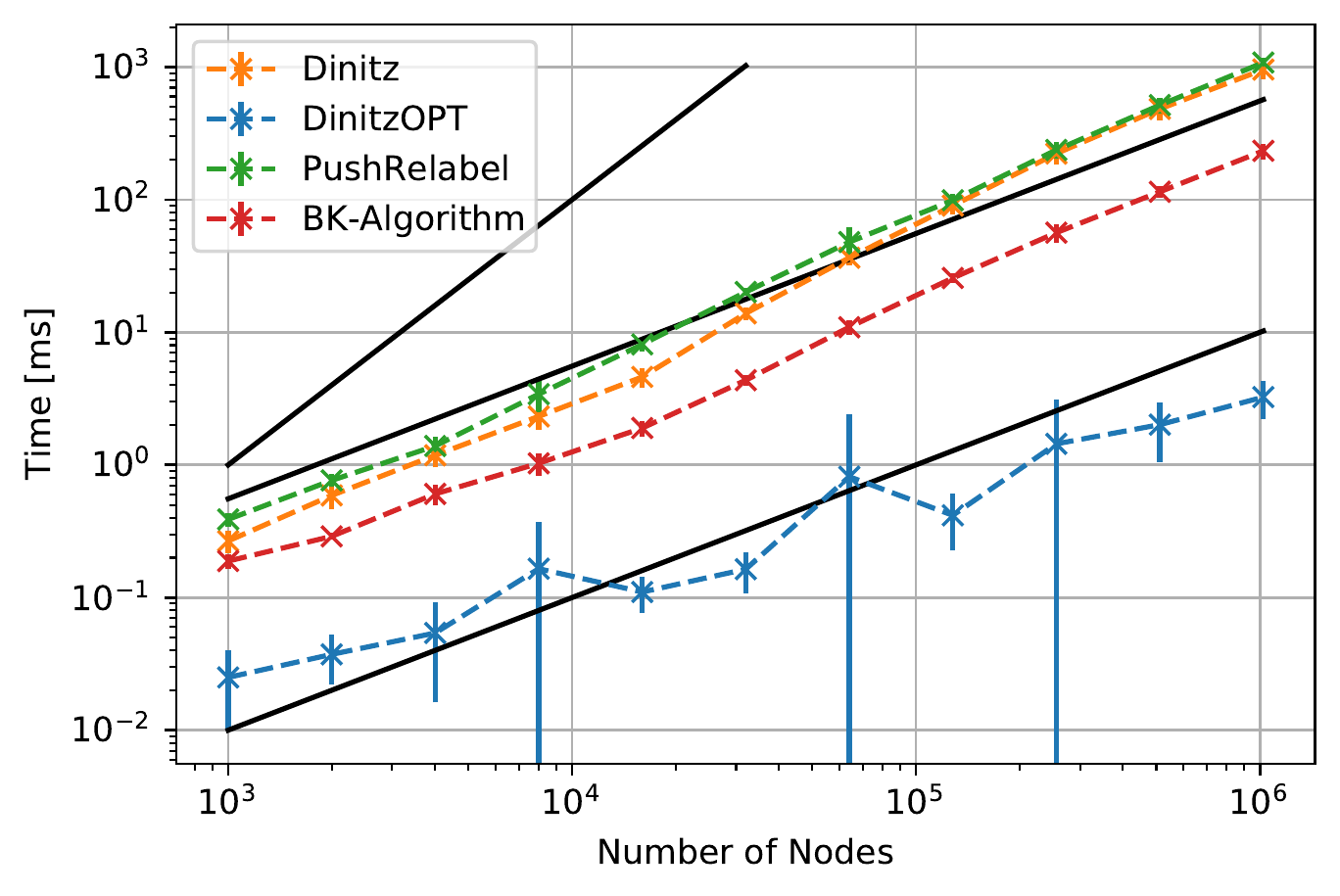}
    \caption{Runtime scaling of flow algorithms. The plot shows the average time per flow over multiple GIRGs and terminal pairs. Two linear and a quadratic function were added for reference.}
    \label{fig:eval-scaling}
\end{figure}

\begin{figure}[t]
    \centering
    \includegraphics[width=0.9\columnwidth]{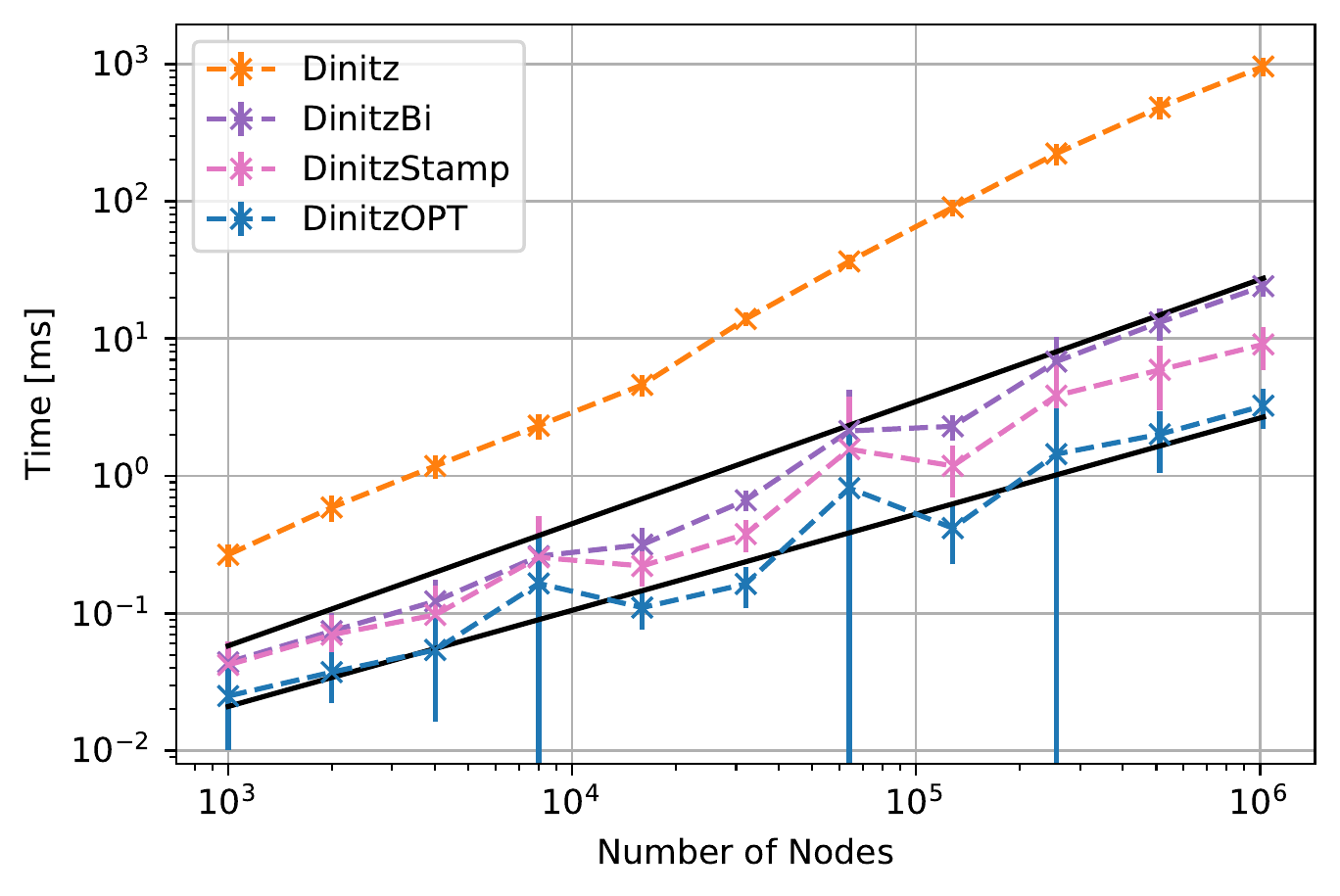}
    \caption{Scaling of Dinitz variants. This plot differs from Figure~\ref{fig:eval-scaling} only in the set of displayed algorithms.}
    \label{fig:eval-dinscaling}
\end{figure}

\paragraph{Scaling.}
We perform additional experiments to analyze the scaling behavior of the algorithms.
Since real networks are scarce and fixed in size, we generate synthetic networks to gradually increase the size while keeping the relevant structural properties fixed.
Geometric Inhomogeneous Random Graphs (GIRGs)~\cite{Bringmann2019}, a generalization of Hyperbolic Random Graphs~\cite{Krioukov2010}, are a scale-free generative network model that captures many properties of real-world networks.
The efficient generator~\cite{Blaesius2019} allows us to benchmark our algorithms on differently sized networks with similar structure.
Figure~\ref{fig:eval-scaling} and Figure~\ref{fig:eval-dinscaling} show the results.

We measure the run time over a series of GIRGs with the number of nodes growing exponentially from \SI{1000} to \SI{1024000}{} with 10 iterations each.
In each iteration, we sample a new random graph with average degree 10, power-law exponent 2.8, dimension 1, and temperature~0.
The run time for each algorithm is then averaged over 10 uniform random pairs of vertices with degree between 10 and 20.
Standard deviation is shown as error bars.
The lower half of the symmetric error bars seems longer due to the log-axis.
We add five functions in black as reference:
a quadratic and two linear functions in Figure~\ref{fig:eval-scaling} and $n^{0.88}$ and $n^{0.7}$ in Figure~\ref{fig:eval-dinscaling}.

Dinitz, Push-Relabel and BK show a near-linear running time. 
Compared to the linear functions in Figure~\ref{fig:eval-scaling}, Dinitz and Push-Relabel seem to scale slightly worse than linear, while DinitzOPT scales better than linear.
In a construction with super-sink and super-source, a similar scaling was observed for Push-Relabel on the Yahoo Instant Messenger graph~\cite{Lang2004a}.
We added the function $n^{0.88}$ to Figure~\ref{fig:eval-dinscaling}, because it is the theoretical upper bound for the bidirectional search on hyperbolic random graphs with the chosen power-law exponent~\cite{Blaesius2018}. 
Also it appears to be a good estimate for Dinitz running time with just the first optimization of bidirectional search (DinitzBi).
It was previously observed that bidirectional search on hyperbolic random graphs with the chosen parameters usually scales like $n^{0.7}$~\cite{Blaesius2018}, which fits the run time of DinitzOPT in our experiments.

Finally, the standard deviation and shape of the curve confirms our claim that the run time of DinitzOPT is more sensitive to the graph structure.
In fact, comparison with our intermediate versions of Dinitz shows that, while bidirectional search improved run time the most, each successive optimization increased the sensitivity to the graph structure.

\subsection{Optimizations in Detail.}
\label{sec:eval-detail}
%     Question 2: Discuss progression DinicsVanilla to DinicsOPT

\newcommand{\DinitzVanilla}{Dinitz\xspace}
\newcommand{\DinitzBi}{DinitzBi}
\newcommand{\DinitzReset}{DinitzReset}
\newcommand{\DinitzStamp}{Dinitz\-Stamp\xspace}
\newcommand{\DinitzOPT}{DinitzOPT\xspace}

\begin{table*}
  \definecolor{tablegray}{gray}{0.9}
  \definecolor{othergray}{gray}{0.6}
  \rowcolors{1}{white}{tablegray}
  \centering
  \caption{
Total run times and search space of visited edges for the five intermediate versions of our Dinitz implementation during the computation of 1000 flows in \texttt{as-skitter}.
Terminals are chosen like \emph{low} pairs in Figure~\ref{fig:eval-runtime}.
The first seven columns show times in seconds accumulated over all flow computations.
BUILD is the construction of the residual network that is reused for all flow computations, 
RESET means clearing flow on edges between computations, 
INIT includes initialization of distances and counters per round, 
BFS and DFS refer to the respective subroutines, 
FLOW is the summed time during flow computations (sum of BFS, DFS, INIT), 
and TOTAL is the run time of the whole application including reading the graph from file.
The last three columns contain the search space relative to the number of edges in the graph in percent.
Search space columns for BFS and DFS are per round, while the FLOW column lists the search space per flow, e.g.,
\DinitzVanilla visits on average 65.66\% of all edges per BFS and every edge is visited about 5.58 times on overage in one flow computation.
  }
  \begin{tabular}{l c c c c c c c c c c}
    \toprule
            &       &       & \multicolumn{3}{c}{MaxFlow} & & & \multicolumn{3}{c}{Search Space [\%]} \\
    \cmidrule{4-6}\cmidrule{9-11}
            & BUILD & RESET & INIT  & BFS   & DFS   & FLOW  & TOTAL & BFS   & DFS   & FLOW  \\
    \midrule
    \DinitzVanilla  & 0.50  & 56.79 & 14.87 & 405.46& 426.80& 847.13& 904.85& 65.66 & 63.64 & 558.04 \\
    %Bidir   & 0.55  & 58.15 & 21.02 & 2.78  & 8.94  & 32.73 & 91.82 & \multicolumn{3}{c}{} \\
    %Reset   & 0.50  & |     & 20.73 & 2.47  & 8.01  & 31.20 & 32.06 & \cellcolor{tablegray}0.26  & \cellcolor{tablegray}1.87  & \cellcolor{tablegray}8.38 \\
    %Stamps  & 0.55  & |     & |     & 2.51  & 10.30 & 12.81 & 13.72 & \multicolumn{3}{c}{} \\
    \DinitzBi       & 0.55  & 58.15 & 21.02 & 2.78  & 8.94  & 32.73 & 91.82 & 0.26  & 1.87  & 8.38 \\
    \DinitzReset    & 0.50  & |     & 20.73 & 2.47  & 8.01  & 31.20 & 32.06 & \textcolor{othergray}{0.26}  & \textcolor{othergray}{1.87}  & \textcolor{othergray}{8.38} \\
    \DinitzStamp      & 0.55  & |     & |     & 2.51  & 10.30 & 12.81 & 13.72 & \textcolor{othergray}{0.26}  & \textcolor{othergray}{1.87}  & \textcolor{othergray}{8.38} \\
    \DinitzOPT     & 0.55  & |     & |     & 2.40  & 1.06  & 3.46  & 4.22  & 0.26  & 0.20  & 2.03 \\
    \bottomrule
  \end{tabular}
  \label{tab:eval-profiler}
\end{table*}

In this section we evaluate the performance impact of the changes discussed in Section~\ref{sec:opt}.
We present a search space analysis and in-depth profiler results\footnote{We used the Intel VTune profiler.}.
In addition to the unmodified Dinitz, we consider four incrementally more optimized versions of the algorithm:
\DinitzBi, \DinitzReset, \DinitzStamp, and \DinitzOPT.
Each algorithm corresponds to adding one optimization to the previous ones.

\paragraph*{Experimental Setup.}

% order
All optimizations can be applied in any order and combination. 
Instead of considering all combinations, we individually add them in a specific order, such that the next change always tackles a performance bottleneck.
In fact, additional benchmarks reveal that the current optimization speeds up the computation more than enabling all other remaining changes together.

\begin{figure}
\centering
\includegraphics[width=\columnwidth]{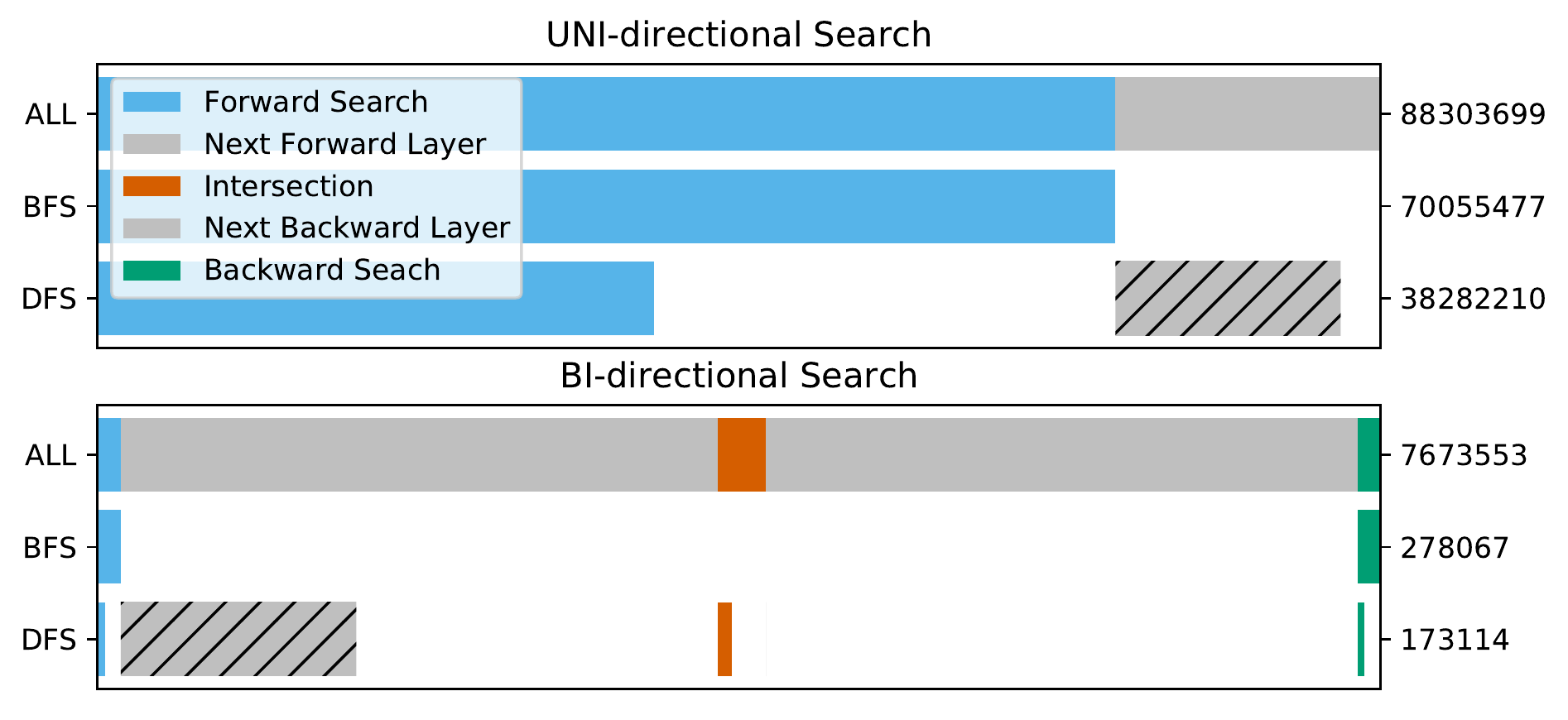}
\caption{Average number of edges visited per flow computation for the terminal pairs used in Table~\ref{tab:eval-profiler}, partitioned as in Figure~\ref{fig:opt-space}.
\emph{Forward/Backward Search} represent the edges explored by the respective search. 
\emph{Next Forward/Backward Layer} denote the edges that would be explored in the next step of the BFS.
Edges in the \emph{Intersection} originate from vertices in both upcoming BFS-layers. 
The BFS and DFS bars show the edges that are actually visited by the algorithm.
%When counting edges for DFS we exclude the path that is augmented to maintain the total space, visualized by the ALL bar, as an upper bound for the DFS and BFS search space.
The shaded area indicates the edges skipped by our last optimization (from \DinitzStamp to \DinitzOPT in Table~\ref{tab:eval-profiler}) and is excluded in the sum on the right.}
\label{fig:eval-space-total}
\end{figure}

% stats, graph, pairs ...
The experiments and benchmarks in this section consider 1000 uniform random terminal pairs close to the average degree on the \texttt{as-skitter} instance.
The average distance between source and sink in the initial network is 4.2.
The average number of rounds until a maximum flow is found is 4.8, where the last round runs only the BFS to verify that no augmenting path exists.
Only counting rounds before the last round, 2.9 units of flow are found on average per round.
Out of the 1000 cuts, 882 have value equal to the degree of the smaller terminal.
Table~\ref{tab:eval-profiler} shows profiler results and search space for Dinitz and the optimized versions of the algorithm.
Additionally, Figure~\ref{fig:eval-space-total} compares the search space with and without bidirectional search.

\paragraph*{Bidirectional Search.}

\DinitzVanilla takes 15 minutes to compute the \SI{1000}{} flows and the search space per flow is more than five times the number of edges on average.
Almost all of that time is spent in BFS or DFS.
The bidirectional Dinitz reduces the flow-time from 14 minutes to 30 seconds, an improvement by a factor of~25.

The search space is reduced by factors of 252 for BFS, 34 for DFS, and 67 per flow.
It is interesting to note, that the search space of BFS during the last round of each flow changes even more.
In this round the BFS will find no s-t path.
The bidirectional search visits 39 edges on average, while the normal breadth-fist-search visits 44\% of the graph.
This not only emphasizes that the cuts are close around one terminal, but also shows that the bidirectional search heavily exploits this structure.

The run time does not fully reflect this drastic reduction in search space, because DFS and BFS no longer dominate the flow computation.
The initialization time per round increased by 50\%, which can be explained by the additional distance label per node to store the distance to the sink (now 3 ints instead of 2).
Although the initialization is a simple linear operation in the number of nodes, it takes twice as long as BFS and DFS combined.
Actually, the performance of initialization heavily depends on the data layout.
We decided to store node data interleaved instead of in separate buffers.
This data layout reduces memory loads and facilitates cache locality because all data for one node is fetched at once.
On the other hand, the choice hinders efficient initialization with SIMD instructions.

The real bottleneck, however, is to reset the flow values between computations.
RESET takes almost a full minute which is twice as long as computing the flows.

\paragraph*{Reset flow between computations.}

%For vision problems, initialization times already compete with max-flow computation times \cite{Verma2012}.

Between flow computations, the residual capacity of all edges has to be reset before another flow can be found.
After changing the BFS to a bidirectional search, resetting the flow on all edges between computations dominates the run time.
To reduce the time of our benchmarks, and to make the code more efficient in situations where multiple flows are computed in the same network, we address this bottleneck.
Instead of explicitly resetting flow values for all edges, we remember the edges that contain flow and reset only those.
The number of edges with positive flow is typically very small in comparison to the whole network.
Additionally, edges that contain flow are visited during the algorithm anyway.
By storing changed edges during DFS, reset flow takes at most as long as augmenting the flow in the first place.
In fact, the time to reset flow is so low, it is not detected by the profiler.
%Note that the edges that are reset here are exactly the edges that are not counted for DFS in Figure~\ref{fig:eval-space-total}.
This change is not mentioned in Section~\ref{sec:opt} because it does not speed up a single flow computation.

This change completely eliminates the time for RESET, while other operations are not affected.
The total time to compute all 1000 flows is thus three times lower with the flow computation making up for almost all spent time.
The slowest part of the flow computation itself is still the initialization with 21 of the 31 seconds.

\paragraph*{Time Stamps.}

The distance labels and counters per node are initialized each round.
Using time stamps eliminates the need for initialization completely while adding a small overhead to DFS.
The flow computation gets 2.4 times faster with 13 seconds instead of 31.
After introducing the time stamps, the DFS is the new bottleneck and makes up for about 80\% of flow time.

\paragraph*{Skip Next Forward Layer.}

As discussed in Section~\ref{sec:opt}, this change prevents the DFS from visiting vertices beyond the last layer of the forward search that are not also seen by the backwards search.
In Figure~\ref{fig:eval-space-total} the skipped part is shaded. 
This optimization reduces the average search space for DFS during one round from almost 2\% of all edges to just 0.2\%.
The improvement in search space is reflected by the profiler results. 
DFS is sped up from 10 seconds to just one second, which is faster than the BFS.
The resulting time to compute all 1000 flows is 3.46 seconds, which is only 7 times slower than building the adjacency list in the beginning.
In total, the time to compute the flows with the optimized Dinitz is 245 times faster than the unmodified Dinitz.

\paragraph*{Misc.}

Since the BFS is the slowest part of the final algorithm, we add another low-level optimization for undirected networks.
Line-by-line load analysis shows that more time is spent during the backward search than the forward search.
The backward search from the sink has to consider incoming instead of outgoing edges but our implementation only maintains an adjacency list of outgoing edges.
However, for each incoming edge there is an outgoing twin edge with a reference to the incoming edge.
This reference is used to determine the residual capacity of the incoming edge to check if the incoming edge is part of the residual network.

We can save a memory lookup in the hot code of the algorithm, by determining the residual capacity of the incoming edge without loading it into memory.
The residual capacity of an edge is obtained by subtracting the flow from the capacity.
In undirected networks the capacity of an edge is the same as that of its twin.
Additionally, consistency of flow links the flow of both edges.
Thus we can compute the residual capacity of incoming edges by looking only at the outgoing edges.
The change improves performance by 20 to 40 percent in undirected networks.

Note that a similar optimization is possible for directed networks: one can cache the capacity of the back edge in each twin.
This concept is known and was applied in previous flow implementations\footnote{\url{https://github.com/Zagrosss/maxflow}}, however we only use the optimization for undirected networks.

\subsection{Gomory-Hu Trees.}
\label{sec:eval-gh}

\begin{figure*}
\includegraphics[width=\textwidth]{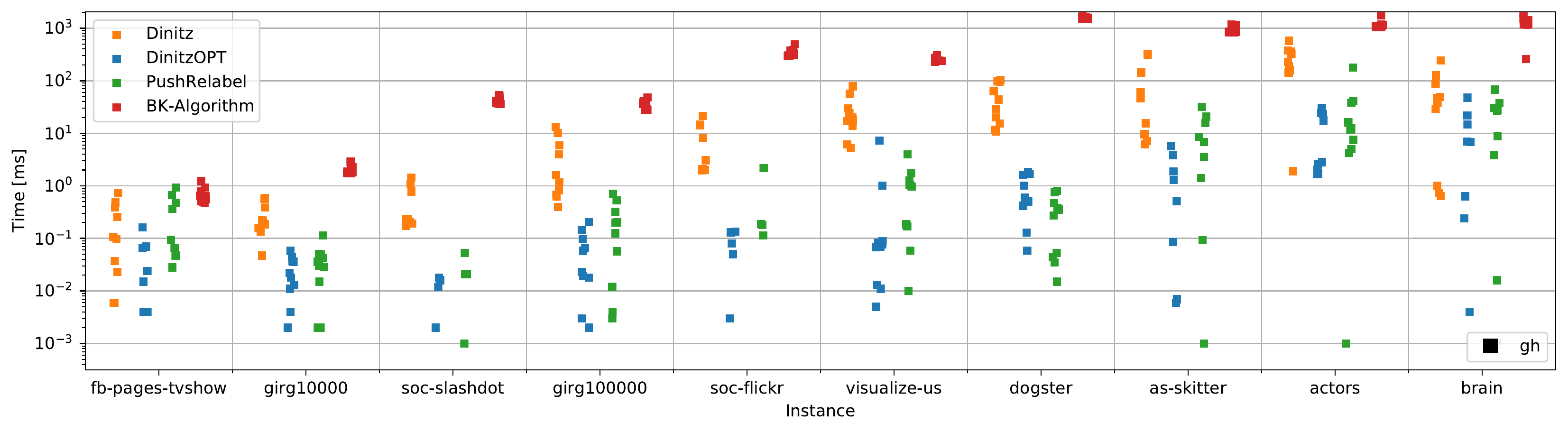}
\caption{Runtime comparison of flow computations. The 10 terminal pairs per instance are uniformly chosen out of the $n-1$ cuts required by Gusfield's algorithm.}
\label{fig:eval-runtime-gh}
\end{figure*}

In the last sections we observed that heterogeneous network structure yields easy flow problems that can be solved significantly faster than the construction of the adjacency list.
This performance becomes important in applications that require multiple flows to be found in the same network.
Gomory-Hu trees~\cite{Gomory1961} fit this setting and have applications in graph clustering~\cite{Flake2004}.
% what is GH tree
A Gomory-Hu tree (GH-tree) of a network is a weighted tree on the same set of vertices that preserves minimum cuts, 
i.e.,~each minimum cut between any two vertices $s$ and $t$ in the tree is also a minimum $s$-$t$ cut in the original network.
Thus, they compactly represent $s$-$t$ cuts for all vertex pairs of a graph.
For the construction of a GH-tree, there exists a very simple algorithm by Gusfield~\cite{Gusfield1990} that requires $n-1$ cut-oracle calls in the original graph.

% intro
In this section we evaluate the performance of max-flow algorithms for the construction of a Gomory-Hu tree in heterogeneous networks.
We will see that the terminal pairs required for Gusfield's algorithm yield easier flow problems than uniform random pairs.
DinitzOPT is able to make use of this easy structure to achieve surprisingly low run times, so is Push-Relabel when only considering the computation of the flow value.
However, we find that the need to extract the source side of the cut hinders Push-Relabel to benefit from this performance.

\paragraph{Flow Computation on Gusfield Pairs.}

% figure gh-pairs
Figure~\ref{fig:eval-runtime-gh} shows the same networks and algorithms as in Figure~\ref{fig:eval-runtime} but with terminal 
pairs sampled out of the $n-1$ flow computations needed by Gusfield's algorithm.
The run times for all algorithms except the BK-Algorithm have high variance and are spread over up to four orders of magnitude for the larger instances.
Although results for different terminal pairs vary greatly, BK seems to be the slowest algorithm followed by Dinitz.
DinitzOPT and PR have comparable but significantly lower run times than the other algorithms.
For example, 6 out of the 10 \emph{gh} pairs measured for the \texttt{soc-slashdot} instance are solved by DinitzOPT and Push-Relabel faster than one microsecond which is the precision of our measurements.
This suggests, that these algorithms are more sensitive to the varying difficulty of the flow computations for \emph{gh} pairs.
Our speedup over the Push-Relabel algorithm on \emph{gh} pairs is not as pronounced as for the random pairs in Section~\ref{sec:eval-runtime}.
On the \texttt{dogster} instance PR is even faster than DinitzOPT.

% PR and OPT have 0us on the same 6/10 gh pairs on soc-slashdot
To further investigate why \emph{gh} pairs are this easy to solve, we analyze a complete run of all pairs needed by Gusfield's algorithm on the \texttt{soc-slashdot} instance.
In Gusfield's algorithm each vertex is the source once, thus the average degree of the source is the average degree of the graph (10.24).
In contrast, the average degree of the sink is ca.~1500, which hinders the benefit of bidirectional search.
Uni-directional Dinitz slows down by a factor of 15 when computing the flows with switched terminals.
%DINICS front and backwards (just flow, no gusfield stuff) in ms
%forward 66'963
%backward 899'945
The average distance between two vertices in the original network is 4.16, but interestingly here the average distance from source to sink is 1.78.
Out of the \SI{70}{\kilo\relax} flow computations, \SI{56}{\kilo\relax} are trivial cuts around one terminal.
Computing a flow for a single s-t pair takes 2.76 rounds on average with the last round only to confirm that the flow is optimal. 
Before the last round on average 5.56 flow is being found per round.

% pairs vs complete run
DinitzOPT and Push-Relabel are both extremely fast on \emph{gh} pairs.
DinitzOPT takes 2.5 seconds to compute all $n=\SI{70}{\kilo\relax}$ required flows, while PR needs 5 seconds.
To obtain the 5 seconds for PR we exclusively measured the preflow computation, but PR is not limited by the time to compute the preflow.
Actually, the entire computation of the Gomory-Hu tree on the \texttt{soc-slashdot} instance takes 12 minutes with Push-Relabel and 2.6 seconds with DinitzOPT.
Instead of being caused by the Gusfield logic --- which actually makes up less than 3\% of the run time when using DinitzOPT as oracle --- the bottleneck when using PR as a cut oracle is not the flow computation, but initialization and extracting the cut.
The drastic difference in run time is in part due to the optimizations we added to DinitzOPT to reduce time between flow computations, while the Push-Relabel implementation recreates the auxiliary data structures, except the adjacency list, before each flow.
However, in the following we will see that a large amount of Push-Relabels run time is actually necessary to extract the cuts for Gusfield's algorithm.

\paragraph{Measuring Gusfield's Algorithm.}

% new try

In Gusfield's algorithm we have to iterate over all vertices in the source-side of the cut.
Extracting these with the Push-Relabel algorithm is slower than with Dinitz.
We outline the three approaches to extract the cut with Push-Relabel and show that each has major drawbacks.

The PR algorithm is executed in two stages.
The first stage computes a preflow and the second stage converts the preflow into a flow.
%A preflow is a flow that violates the conservation of flow constraint and allows an excess of flow in a vertex.
Often, computing a preflow is sufficient, because one obtains the value of the max-flow/min-cut and can determine a cut by finding all sink-reaching vertices in the residual network.
Since Gusfield requires the source-side of the cut, the complement of the found set of vertices can be used.
This approach is computationally expensive, because of the high sink degree.

Given a max-flow, one partition of a min-cut can also be identified by reachability from the source in the residual network.
Since the source usually has a smaller degree during Gusfield's algorithm, the source-side of the cut is small. 
This approach is efficient and can be used for Dinitz.
However, for PR it requires the preflow to be converted into a flow.
Asymptotically, the first stage (preflow) dominates the second (convert) stage, but in practice this is not always the case.
In the paper that proposed the current PR implementation~\cite{Cherkassky1997} the authors experiment with different implementations of the conversion 
and find a method whose "running time [\dots] is a small fraction of the running time of the first stage".
Other works find that 95\% of time is spent in stage one~\cite{Derigs1989}.
Our experiments in Section~\ref{sec:eval-runtime} are in line with these findings and thus only the time for the first stage of PR is shown there.
However, Gusfield pairs pose easily solvable flow instances due to the low distance between source and sink.
Thus, the simplicity causes the second stage of PR to dominate the first one.

The drawbacks of the two previous approaches can be avoided in undirected networks by computing the preflow from sink to source.
Without preflow-conversion, a cut can then be extracted by determining the vertices that can reach the original source in the residual network.
The drawback of this method is that the preflow computation slows down massively.

In short, the three approaches to extract the source-side of a min-cut with the Push-Relabel algorithm are:
\begin{description}
\item[Convert.] Compute a preflow from the source, convert it into a flow, then run BFS from the source.
\item[T-Side.] Compute a preflow from the source, run BFS backwards from the sink, then take complement.
\item[Swap.] Compute a preflow from sink to source, then run BFS backwards from the source.
%\item[Useless] \TODO{For completeness but useless: swap s,t; convert; reach from s; invert.}
\end{description}

\begin{figure}
\centering
\includegraphics[width=0.5\textwidth]{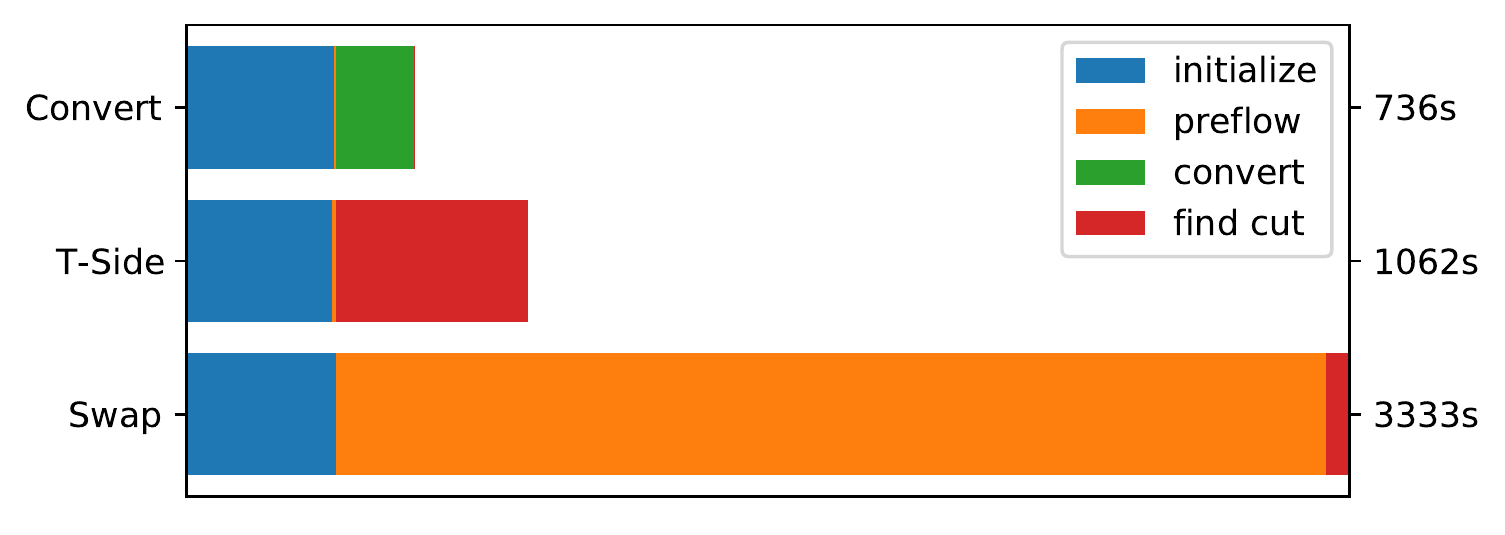}
\caption{Distribution of spent time during Gusfield's algorithm on the \texttt{soc-slashdot} instance with three approaches to use the Push-Relabel algorithm as a min-cut oracle.
We split the measurements into initialization, preflow, conversion, and cut identification.
The time overhead for measurement, logging, and the logic of Gusfield's algorithm is included in the numbers on the right but excluded in the bars.
}
\label{fig:eval-gh}
\end{figure}

Figure~\ref{fig:eval-gh} shows the distribution of run time using these approaches to run Gusfields algorithm on the \texttt{soc-slashdot} instance. 
The \emph{convert} approach is the fastest with just above 12 minutes followed by \emph{T-side} with 18 minutes and \emph{swap} with almost an hour.
The initialization time provides a reference, because it is approximately 7 minutes for all approaches.
We note here that the initialization for PR creates some Boost related data and performs an operation linear in the number of edges.

% convert
We see that the flow computation is actually really fast for \emph{convert}.
It takes only about 5 seconds of these 12 minutes.
The initialization dominates this time and the conversion is also far slower than the flow itself.

% tside
Surprisingly, the flow takes twice as long for \emph{T-side} than for \emph{convert}, although only the way to identify the cut was changed.
This is, because we find other min-cuts and thus obtain a different GH-tree while processing different terminal pairs.
We also implemented the \emph{T-side} approach for DinitzOPT to verify the correctness of the computed cuts and trees.
Interestingly running this takes 4.5 minutes which is a factor 100 slower than identifying the cut via the source-side for DinitzOPT.
Similarly for PR, we observe that the cut identification, which was almost unnoticeable for \emph{convert}, makes up most of the computation time for \emph{T-side}.

% swap
Lastly, the \emph{swap} approach takes more than 4 times as long as the \emph{convert} approach.
As the degree of the sink is significantly larger than the source, the flow computation slows down massively.
It goes from 5 seconds to 47 minutes.
Recall that the unmodified Dinitz slows down by a factor of 15 when switching source and sink.
%\TODO{why is cut identification not as fast as for convert; cache?}

In conclusion, all three methods perform significantly worse than DinitzOPT, not because PR flow computations are slow, but because the initialization and cut identification 
already take orders of magnitude longer than the complete process with DinitzOPT.
Both methods to avoid the four minutes run time of stage two of the Push-Relabel algorithm imply even worse performance cost;
either due to a breadth-first search that has to traverse almost the whole graph (T-side) or due to significantly slower preflow computations (Swap).

\subsection{Other Types of Networks.}
\label{sec:eval-other}

\newcommand{\Erdos}{Erd\H{o}s-R{\'e}nyi}

\begin{figure*}
\centering
\includegraphics[width=\textwidth]{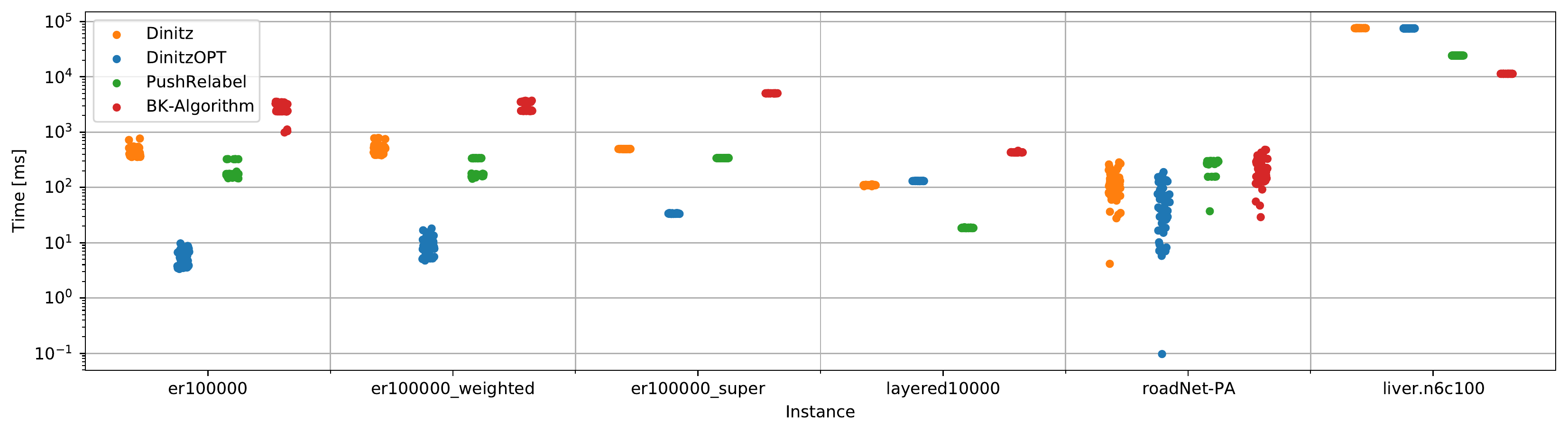}
\caption{Run time of max-flow computations for various networks.
Each point corresponds to one $s$-$t$ flow. 
%For \texttt{er*} and \texttt{roadNet-PA} the terminals were chosen uniformly at random, while \emph{layered} and \emph{liver} have fixed terminals.
%The algorithms completely rebuild their internal data structures between computations, which is excluded in the measurement.
For each instance we computed 50 $s$-$t$ flows.
The instances \texttt{er100000\_super}, \texttt{layered10000}, and \texttt{liver.n6c100} have designated terminals. 
For \texttt{er100000}, \texttt{er100000\_weighted}, and \texttt{roadNet-PA} terminals are chosen uniformly at random.
Unlike the experiments in Section~\ref{sec:eval-runtime}, the algorithms rebuild their internal data structures including the adjacency list before each flow computation.
This was necessary to prevent the BK-algorithm from reusing search-trees, which makes the instances with given terminal pairs trivial after the first run.
}
\label{fig:eval-runtime-other}
\end{figure*}

After evaluating the performance on heterogeneous networks we extend our experiments to networks of different structures.
We consider the following networks:
an Erd\H{o}s-R{\'e}nyi random graph~\cite{Erdos1959} (\texttt{er100000}),
an Erd\H{o}s-R{\'e}nyi random graph with uniform random weights in $[500,10000]$ (\texttt{er100000\_weighted}),
an Erd\H{o}s-R{\'e}nyi random graph with super terminals (\texttt{er100000\_super}),
a generated layered network~\cite{Ahuja1997} (\texttt{layered10000}),
the road network of Pennsylvania (\texttt{roadNet-PA}),
and a liver CT scan as a regular 6-connected grid (\texttt{liver.n6c100}).  
Further details regarding the datasets can be found in Section~\ref{sec:data}.

Figure~\ref{fig:eval-runtime-other} shows the performance of the flow algorithms on these instances.
The performance on the \Erdos~graphs is similar to our results for heterogeneous networks;
the BK-algorithm is the slowest, followed by Dinitz, Push-Relabel, and DinitzOPT in this order.
Note that a running time close to $O(\sqrt n)$ was shown for bidirectional search on \Erdos~random graphs~\cite{Borassi2016}.
Neither weights nor higher-degree terminals change how the algorithms compare among each other.

The layered network, which is specifically constructed to produce a computationally difficult flow instance~\cite{Ahuja1997}, is indeed more difficult than the others.
In the layered network, Push-Relabel is at least five times faster than Dinitz.
DinitzOPT is 10-20\% slower than Dinitz.
After all, our optimizations trade a small overhead during flow computation for the possibility of sublinear running time on particularly easy instances.

For the road network, the choice of the algorithm does not matter as much as for the other instances.
The choice of the terminal pair, however, affects the performance immensely.
With a diameter of almost 800 and a very homogeneous degree distribution, the uniform random choice of terminal pairs produces problems of varying difficulty.
Dinitz, BK, and DinitzOPT capitalize on the easier pairs, while Push-Relabel shows less variance between pairs.

Lastly, the liver scan produces different results than previous instances.
The BK-algorithm was specifically designed for this kind of network structure and application.
Unsurprisingly, the BK-algorithm performs best, followed by Push-Relabel, Dinitz, and DinitzOPT.

%\subsection*{Clustering}

%We computed a clustering according to this paper~\cite{Flake2004}. 
%This involves non-integer capacities, but looks nice.
%Result is shown in Figure~\ref{fig:eval-clustering}.
%Do you think that this is interesting Thomas?
%It does not really fit in this paper, does it?

%\begin{figure*}
%\centering
%\includegraphics[width=0.95\textwidth]{eval_clustering.pdf}
%\caption{A nice picture, don't you think?}
%\label{fig:eval-clustering}
%\end{figure*}

\section{Conclusion}

We presented a modified version of Dinitz's algorithm with greatly improved run time and search space on real-world and generated scale-free networks.
The scaling behavior appears to be sublinear, which matches previous theoretical and empirical observations about the running time of balanced bidirectional search in scale-free random networks.
While these theoretical bounds apply during the first round of our algorithm, it is still unknown whether the analysis can be extended to account for the changes in the residual network.
Our experiments, however, indicate that the search space remains small in subsequent rounds.

We observe that that the low diameter and heterogeneous degree distribution leads to small and unbalanced cuts that our algorithm finds very efficiently.
The flow computations required to compute a Gomory-Hu tree are even easier, making usually insignificant parts of the tested algorithms be a bottleneck.
For example, the preflow conversion leads to Push-Relabel being greatly outperformed by our algorithm in this setting.

Our results on other types of instances show that their structural properties play a huge role when comparing flow algorithms.
It is not surprising that our algorithm is outperformed by the BK-algorithm, which was specifically designed for vision problems, on \texttt{liver.n6c100}.
Moreover, the experiments on the artificial \texttt{layered1000} instance indicate that Push-Relabel is more robust regarding hard instances.
On scale-free networks, however, we drastically improve performance over existing algorithms.

\clearpage

\bibliography{ms}

\clearpage

\appendix

\section{Appendix}

\subsection{Implementation Details}
\label{sec:impl}

Experiments were done on a Dell XPS 15 9570 Laptop with an Intel Core i7-8750H CPU.

\paragraph*{BK-Algorithm.}
As a BK implementation we use the one that was written for the original paper \cite{Boykov2004} provided on the web page of Vladimir Kolmogorov\footnote{\url{http://pub.ist.ac.at/~vnk/software.html}}.
For each $s$-$t$ flow we add edges with huge capacity between $s,t$ and the virtual terminals. 
After the flow is computed, we remove these edges again.
This $O(1)$ work is included in time measurements.
We apply the \emph{reuse trees} feature and mark the changed terminals between flow computations accordingly.
Internal memory is allocated on network construction and not per flow.
There is a BK implementation available in Boost\footnote{\url{https://www.boost.org/doc/libs/1_72_0/libs/graph/doc/boykov_kolmogorov_max_flow.html}}.
We found the original one easier to use, because its interface is tailored towards multiple flow computations and provides easy and efficient access to the found cut.

\paragraph*{Push-Relabel}
The original implementation, used for example in \cite{Verma2012}, is no longer available\footnote{was \url{http://www.avglab.com/andrew/soft.html}}.
We use the C++ version of the original implementation provided in Boost\footnote{\url{https://www.boost.org/doc/libs/1_72_0/libs/graph/doc/push_relabel_max_flow.html}}.
The Boost version is mostly the same code (up to same variable names) ported to C++, but is data structure agnostic.
Therefore, we had to reimplemented the linearised adjacency list data structure used in the original implementation.

\paragraph*{Dinitz and DinitzOPT.}
Our implementation is based on a version of Dinitz that is usually used in programming competitions\footnote{\url{https://cp-algorithms.com/graph/dinic.html}}.
We changed the graph representation to a linear adjacency list of outgoing edges.
Edges are sorted by originating vertex in linear time. 
Each node stores a range of edges into this list.
This is the same structure used for the Push-Relabel implementation.
Performance-wise, the data structure significantly reduces the time to build large networks, but flow time remains the same.
We use an array of size $n$ as a queue, because during BFS each vertex is pushed at most once.
We allocate memory for distance labels, counter, and the queue in advance when the network is built instead of per flow.
In the unidirectional BFS, one could break when the sink is encountered but we finish the current layer for the purpose of measuring search space.

\paragraph*{Undirected Networks.}
We support flow for undirected networks. 
A simple way to do this, is to represent each undirected edge as two directed edges, which was done for Push-Relabel.
However, each directed edge already implies two edges in the residual network: one with the given capacity, and a reversed twin edge with no capacity.
To avoid storing four times the amount of edges, the twin edge can be used to implement undirected flow.
By giving the twin edge the same capacity as its counterpart, the exact same implementation can be used for undirected as well as directed networks.

\paragraph*{Non-integer capacities.}
We use 64-bit floating point numbers instead of integers to represent flow values and capacities, because some applications use non-integer capacities. 
The same implementation can be used but requires more memory and additional checks to handle floating point imprecision.
We applied this to Dinitz, PR, and BK and observed a performance drop of approximately 10\% for all algorithms.
Note that the range in which 64-bit floats exactly represent integral numbers even exceeds the range of 32-bit integers.
Precision issues are cause by the \emph{infinity} capacity edges introduced for BK.
To resolve this, the representation of infinity on these edges must be chosen according to the range of capacities.

\begin{table*}
  \definecolor{tablegray}{gray}{0.9}
  \rowcolors{1}{white}{tablegray}
  \centering
  \caption{Instances used in this paper. The road network was undirected and is converted to directed DIMACS format. In this case, the number of edges refers to the undirected version.}
  \begin{tabular}{l c c r r r l}
    \toprule
    instance & directed & weighted & nodes & edges & avg. degree & source \\
    \midrule
    fb-pages-tvshow     &           &           &   4K &    17K &   8.87 & \href{http://networkrepository.com/fb-pages-tvshow.php}{Network Repository} \\
    girg10000           &           &           &   10K &   60K &  11.99 & generated \\
    soc-slashdot        &           &           &   70K &  360K &  10.24 & \href{http://networkrepository.com/soc-slashdot.php}{Network Repository} \\
    girg100000          &           &           &  100K &  600K &  12.00 & generated \\
    soc-flickr          &           &           &  514K &  3.2M &  12.42 & \href{http://networkrepository.com/soc-flickr.php}{Network Repository} \\
    visualize-us        &           &\checkmark &  594K &  3.2M &  10.92 & \href{http://networkrepository.com/visualize-us.php}{Network Repository} \\
    dogster             &           &           &  427K &  8.5M &  40.03 & \href{http://konect.cc/networks/petster-friendships-dog}{U. Koblenz} \\
    as-skitter          &           &           &  1.7M & 11.1M &  13.08 & \href{https://snap.stanford.edu/data/as-Skitter.html}{U. Stanford} \\
    actors              &           &\checkmark &  382K & 15.0M &  78.69 & \href{http://konect.cc/networks/actor-collaboration}{U. Koblenz} \\
    brain               &           &           &  178K & 15.8M & 176.47 & \href{http://networkrepository.com/bn-human-BNU-1-0025890-session-1.php}{Network Repository} \\
    er100000            &\checkmark &           &  100K &   20M & 199.94 & generated \\
    layered10000        &\checkmark &           &   10K &  100K &   9.96 & generated \\
    roadNet-PA          &(\checkmark)&          &  1.1M &   1.5M&   2.83 & \href{https://snap.stanford.edu/data/roadNet-PA.html}{U. Stanford} \\
    liver.n6c100        &\checkmark &\checkmark &  4.1M &   25M &   6.04 & \href{https://vision.cs.uwaterloo.ca/data/maxflow}{U. Western Ontario} \\
    %clustering          & * & x &    4K &   16K &   7.38 & generated \\
    \bottomrule
  \end{tabular}
  \label{tab:instances}
\end{table*}

\subsection{Data.}
\label{sec:data}

We obtained the datasets from the University of Koblenz (KONECT)~\cite{Kunegis2013}, 
the Network Repository website~\cite{Rossi2015}, as well as the Stanford Network Analysis Project (Snap)~\cite{Leskovec2014}.

Furthermore, we used the GIRG generator by Bl\"asius et al.~\cite{Blaesius2019} mostly with default parameters.
We implemented the ER model and the layered network construction from Ajuja et al.~\cite{Ahuja1997}.
The parameters for ER are $n=100000$ and $p=0.02$.
The parameters for the layered network are taken from the largest instance in their paper (W=71, L=141, d=10).

Lastly, the \texttt{liver.n6c100} instance is from the University of Western Ontario.
It is a regular 3D grid with 170x170x144 nodes, 6 edges per node, capacities up to 100, and a super sink/source.

We converted all instances to a text-based edge list with zero-based indices.
In Section~\ref{sec:eval-other} we use the directed DIMACS format instead.

\end{document}